\documentstyle[epsfig,12pt]{article}  
\newcommand{\beq}{\begin{equation}}
\newcommand{\eeq}{\end{equation}}
\newcommand{\bea}{\vspace{0.25cm}\begin{eqnarray}}
\newcommand{\eea}{\end{eqnarray}}

\newcommand{\ro}{\mbox{{\boldmath
$\rho$}}}

\newcommand{\pb}{{{\bf p}}}

\newcommand{\bb}{{{\bf b}}}

\def\lsim{\mathrel{\rlap{\lower4pt\hbox{\hskip1pt$\sim$}}
    \raise1pt\hbox{$<$}}}         %less than or approx. symbol
\def\gsim{\mathrel{\rlap{\lower4pt\hbox{\hskip1pt$\sim$}}
    \raise1pt\hbox{$>$}}}         %greater than or approx. symbol
\begin{document}
\vspace*{-2cm}
 
\bigskip
%%%%%%%%%%%%%%%%%%%%%%%%%%%%%%%%%%%%%%%%%%%%%%%%%%%%%%%%%%
%%%%%%%%%%%%%%%%%%%%%%%%%%%%%%%%%%%%%%%%%%%%%%%%%%%%%%%%%% 

\begin{center}

\renewcommand{\thefootnote}{\fnsymbol{footnote}}

  {\Large\bf
Nuclear modification factor
for light and heavy flavors within pQCD
and recent data from the LHC
\\
\vspace{.7cm}
  }
\renewcommand{\thefootnote}{\arabic{footnote}}
\medskip
{\large
  B.G.~Zakharov
  \bigskip
  \\
  }
{\it
 L.D.~Landau Institute for Theoretical Physics,
        GSP-1, 117940,\\ Kosygina Str. 2, 117334 Moscow, Russia
\vspace{1.7cm}\\}

  {\bf
  Abstract}
\end{center}
{
\baselineskip=9pt
We examine the flavor dependence 
of the nuclear modification factor $R_{AA}$ in the pQCD
calculations at LHC energies. The computations are performed accounting for 
radiative and collisional 
parton energy loss with running coupling constant.
Our results show that the recent LHC data on the $R_{AA}$ for
charged hadrons, $D$-mesons and non-photonic
electrons agree reasonably with
the pQCD picture of the parton energy loss with
the dominating contribution from the radiative mechanism.
\\
\vspace{.7cm}
}

\noindent{\bf 1.} 
The parton energy loss in the quark-gluon plasma
(QGP) is widely believed to be a source for strong suppression 
of high-$p_{T}$ hadrons in
$AA$-collisions (usually called  the jet quenching) observed at RHIC and LHC.
Understanding the underlying mechanisms of the parton energy loss
is of great importance for application of the jet quenching to probing
the hot QCD matter produced in $AA$-collisions.
In the pQCD picture fast partons lose energy mostly due to 
induced gluon radiation
\cite{BDMPS,LCPI,BSZ,W1,GLV1,AMY}. The effect of collisional energy loss
\cite{Bjorken1}
for the RHIC and LHC conditions is likely to be relatively small 
\cite{Z_Ecoll,Gale}.
Unfortunately, uncertainties in the pQCD-based models
of the jet quenching remain large (mostly due to difficulties in 
modeling multiple gluon emission). 
For the nuclear modification factor
$R_{AA}$ they are  perhaps about a factor two.
Despite this,
it seems relatively safe to
assume that predictions for variation of the $R_{AA}$ should be more robust,
if the parameters are already adjusted to fit some set of experimental data. 

From the point of view of the underlying physics of the jet quenching
it is very interesting to compare 
$R_{AA}$ for light and heavy flavors. It was suggested \cite{DK} that for 
the heavy quarks the dead cone effect should suppress induced 
gluon emission and give rise to an increase of the $R_{AA}$.
However, the observed at RHIC  strong
suppression of the non-photonic electrons 
from the $B/D-$meson decays \cite{PHENIX1_e,STAR_e,PHENIX2_e}
seemed to be in contradiction with this picture.
It may indicate that for RHIC conditions the dead cone
suppression is not very strong or that the radiative mechanism is
not the dominating one at all. 
It stimulated the renewed interest in the collisional energy loss
\cite{Peshier}. Although, by adjusting the coupling
constant one can obtain a sufficiently strong heavy quark suppression
due to the collisional mechanism alone, this scenario does not seem to 
be realistic (at least for $p_{T}\gsim 5-10$ GeV).
Calculations of 
the radiative and collisional energy losses with the same 
$\alpha_{s}$ and the Debye screening  mass performed in \cite{Z_Ecoll}
clearly demonstrate that the collisional loss is relatively small 
for relativistic partons and unlikely 
to change significantly the heavy quark energy loss (see also \cite{Gale}). 

In \cite{DK} the dead cone suppression was estimated from a qualitative
analysis neglecting the quantum finite-size effects.
Calculations of the induced gluon emission from heavy quarks
in a brick of QGP \cite{AZ} within the light-cone path 
integral (LCPI) approach \cite{LCPI}, which treats accurately the mass effects,
demonstrate that at energy $\sim 10-20$ GeV 
for $c$-quark the induced gluon spectrum 
is very similar to that for light quarks
and $\Delta E_{c}\approx \Delta E_{u,d,s}$, and only for for $b$-quark
the gluon emission is suppressed (but not so strongly as predicted by the dead
cone model \cite{DK}).
At high energies ($\gsim 100-200$ GeV) the radiative energy loss
has an anomalous mass dependence with 
$\Delta E_{b}>\Delta E_{c}>\Delta E_{u,d,s}$ due to the quantum 
finite-size effects in radiation of hard gluons \cite{AZ}.
In light of these results we can expect
that the nuclear modification factor for the heavy quark jets 
for RHIC and LHC conditions
should be qualitatively similar to that
for light partons already at $p_{T}\sim 10-20$ GeV. Although accurate simulations and comparison with
experiment are needed to reach definite conclusions.

In the present work we examine the flavor dependence of the
nuclear modification factor 
within the LCPI approach \cite{LCPI} and compare our results with
the latest LHC data on 
the $R_{AA}$ for 
charged hadrons
\cite{CMS_RAAch,ALICE_RAAch},
$D$-mesons \cite{ALICE_RAA_D1,ALICE_RAA_D2} and  
non-photonic electrons \cite{ALICE_RAAe}
in $Pb+Pb$ collisions at $\sqrt{s}=2.76$ TeV.
In evaluating the nuclear modification factor, besides the radiative  
energy loss, we include the collisional one.
Both the radiative and collisional contributions are 
calculated with running $\alpha_{s}$. 
We  account for accurately 
the fluctuations of the parton path lengths in the QGP.
We find that the predicted flavor dependence of the $R_{AA}$ 
agrees reasonably with the LHC data.

\vspace{.2cm}
\noindent{\bf 2.}
We calculate the nuclear modification factor 
employing the method developed in Ref. \cite{RAA08},
to which the interested reader is referred for details. 
Here we just outline the main aspects of the calculations
necessary for understanding of our strategy
and interpretation of the results.

For a given impact parameter $b$
the $R_{AA}$ can be written as
\beq
R_{AA}(b)=\frac{{dN(A+A\rightarrow h+X)}/{d\pb_{T}dy}}
{T_{AA}(b){d\sigma(N+N\rightarrow h+X)}/{d\pb_{T}dy}}\,.
\label{eq:10}
\eeq
Here $\pb_{T}$ is the particle transverse momentum, $y$ is rapidity (we
consider the central region $y=0$), 
$T_{AA}(b)=\int d\ro T_{A}(\ro) T_{A}(\ro-\bb)$, $T_{A}$ is the nucleus 
profile function. The differential yield 
in $AA$-collision can be written in the form 
\beq
\frac{dN(A+A\rightarrow h+X)}{d\pb_{T} dy}=\int d\ro T_{A}(\ro)T_{A}(\ro-\bb)
\frac{d\sigma_{m}(N+N\rightarrow h+X)}{d\pb_{T} dy}\,,\,\,\,
\label{eq:20}
\eeq
\beq
\frac{d\sigma_{m}(N+N\rightarrow h+X)}{d\pb_{T} dy}=
\sum_{i}\int_{0}^{1} \frac{dz}{z^{2}}
D_{h/i}^{m}(z, Q)
\frac{d\sigma(N+N\rightarrow i+X)}{d\pb_{T}^{i} dy}\,.\,\,\,
\label{eq:30}
\eeq
Here $\pb_{T}^{i}=\pb_{T}/z$ is the parton transverse momentum, 
${d\sigma(N+N\rightarrow i+X)}/{d\pb_{T}^{i} dy}$ is the
hard cross section,
$D_{h/i}^{m}$ is the medium-modified fragmentation function (FF)
for transition of a parton $i$ into the observed particle $h$.
For the parton virtuality scale $Q$ we take the parton transverse
momentum $p^{i}_{T}$.

We assume that the induced radiation stage occurs after the 
DGLAP stage which gives the input
parton distribution for the induced gluon emission stage.
It seems reasonable since for
jets with $E\lsim 100$ GeV the typical time scale for the 
DGLAP stage is relatively small ($\lsim 0.3-1$ fm \cite{RAA08}),  and 
in first approximation it is legitimate to neglect interference
of the DGLAP and the induced gluon emission stages.
Symbolically the medium-modified FF 
reads
\beq
D_{h/i}^{m}(Q)\approx D_{h/j}(Q_{0})
\otimes D_{j/k}^{in}\otimes D_{k/i}(Q)\,,
\label{eq:40}
\eeq
where $\otimes$ denotes $z$-convolution, 
$D_{k/i}$ is the ordinary DGLAP FF for $i\to k$ parton transition,
$D_{j/k}^{in}$ is the FF for $j\to k$ parton transition in the QGP
due to induced gluon emission, and 
$D_{h/j}$ describes 
parton hadronization outside of the QGP\footnote{The approximation
(\ref{eq:40}) ignores creation in the QGP of the anomalous jet color states,
which may be important for the baryon $R_{AA}$ \cite{AZ_baryon} 
at not very high $p_{T}$ and the jet structure in the soft region
\cite{W_color-flow}. But it should be reasonable for evaluating the $R_{AA}$ 
for charged hadrons, which is dominated by the charged pions, and the $R_{AA}$
for heavy flavors.}.
In (\ref{eq:40}) $Q_{0}$ is the scale at which
the DGLAP parton showering is stopped. As in \cite{RAA08} we
take $Q_{0}=2$ GeV.

We computed the DGLAP FFs with the help of the PYTHIA event 
generator \cite{PYTHIA}.
The one gluon 
induced spectrum, $dP/dx$, was calculated within the LCPI approach 
\cite{LCPI} employing the method developed  in 
\cite{Z04_RAA}.
The $D_{j/k}^{in}$ has been obtained from $dP/dx$
accounting for  multiple gluon emission 
within Landau's method as in \cite{BDMS_RAA}.
Note that we include the $q\to g$ transition as well, which is 
usually neglected.
For the $D_{h/j}(Q_{0})$ we use the 
KKP \cite{KKP} FFs  for light partons,
and Peterson FF for heavy quarks (with parameters
$\epsilon_{c}=0.06$ and $\epsilon_{b}=0.006$).
For the non-photonic electrons we evaluated the FFs $c\to e$ 
and $b\to e$ treating them as the two-step fragmentations 
$c\to D\to e$ and $b\to B\to e$. 
The distributions $B/D\to e$ were calculated using the CLEO data 
\cite{CLEO_B,CLEO_D} on the electron spectra in the $B/D$-meson decays.
We neglected the $B\to D\to e$ process, which gives a negligible 
contribution \cite{Vogt}.

The hard cross sections were calculated using the LO 
pQCD formula with the CTEQ6 \cite{CTEQ6} parton distribution functions.
To simulate the higher order effects
we take for the virtuality scale in $\alpha_{s}$ the value 
$cQ$ with $c=0.265$ as in the PYTHIA event generator \cite{PYTHIA}.
This prescription allows us to reproduce well the 
$p_{T}$-dependence of the spectra in $pp$-collisions
\footnote{Although we use the LO formula for the heavy quark
cross sections, the $p_{T}$-dependences (and the $c/b$ ratio)
of our cross sections agree well with the more sophisticated 
FONLL calculations \cite{Vogt} (the normalization of the cross
sections is unimportant for the $R_{AA}$ at all).}.
In calculating the $R_{AA}$ we account for the 
nuclear modification of the parton densities
(which leads to some small deviation of $R_{AA}$ from unity even without
parton energy loss) with the help of the 
EKS98 correction \cite{EKS98}.

As in \cite{RAA08}
we take $m_{q}=300$ and $m_{g}=400$ MeV for the light quark and gluon 
quasiparticle masses supported by the analysis of the lattice
data \cite{LH}. For the heavy quarks we take $m_{c}=1.2$ GeV
and $m_{b}=4.75$ GeV. We use 
the Debye mass obtained 
in the lattice calculations \cite{Bielefeld_Md} that
give the ratio $\mu_{D}/T$ slowly decreasing with $T$  
($\mu_{D}/T\approx 3$ at $T\sim 1.5T_{c}$, $\mu_{D}/T\approx 2.4$ at 
$T\sim 4T_{c}$). 

We use the running $\alpha_s$
frozen at some value $\alpha_{s}^{fr}$ at low momenta
(the technical details for incorporating the running $\alpha_s$
can be found in \cite{Z04_RAA}). For gluon emission
in vacuum a reasonable choice is $\alpha_{s}^{fr}\approx 0.7$
\cite{NZ_HERA,DKT}.
The RHIC data on the pion $R_{AA}$ in $Au+Au$ collisions at
$\sqrt{s}=200$ GeV support $\alpha_{s}^{fr}\sim 0.5-0.6$ \cite{RAA08}.
But the analysis \cite{Z_RHIC-ALICE} of the first LHC data on the 
$R_{AA}$ for charged hadrons in $Pb+Pb$ collisions at $\sqrt{s}=2.76$ TeV 
obtained by ALICE \cite{ALICE1} shows that they
agree better with $\alpha_{s}^{fr}\sim 0.4-0.5$. The calculations with
a fixed coupling constant \cite{Gyulassy1,Gyulassy2} also indicate
that it can be smaller at LHC energies.
The reduction of $\alpha_{s}^{fr}$ from RHIC to LHC
is probably a manifestation of the thermal suppression
of $\alpha_{s}$ due to the growth of the initial temperature of the QGP 
at LHC. We will see that the new data from CMS \cite{CMS_RAAch} 
and ALICE \cite{ALICE_RAAch} also support
$\alpha_{s}^{fr}\sim 0.4-0.5$.

We view the collisional energy loss as a perturbation \cite{RAA08},
and account for its effect simply by redefining the initial QGP 
temperature in calculating the radiative FF according to the
condition
\beq
\Delta E_{rad}(T^{\,'}_{0})=\Delta E_{rad}(T_{0})+\Delta E_{col}(T_{0})\,,
\label{eq:50}
\eeq
where
$\Delta E_{rad/col}$ is the radiative/collisional energy loss, $T_{0}$
is the real initial temperature of the QGP, and $T^{\,'}_{0}$ is the 
renormalized temperature. We solve (\ref{eq:50}) in linear approximation
in $T^{\,'3}_{0}-T^{3}_{0}$, which gives 
$T^{\,'3}_{0}=T^{3}_{0}+\Delta E_{col}(T_{0})/[dE_{rad}(T_{0})/dT^{3}_{0}]$.
It was done for each parton trajectory in the QGP
(separately for quarks and gluons).
The collisional energy loss has been evaluated in the 
Bjorken method \cite{Bjorken1}
with an accurate treatment of kinematics of the binary collisions 
(the details can be found
in \cite{Z_Ecoll}).

\vspace{.2cm}
\noindent{\bf 3.} 
We perform the computations for 
Bjorken 1+1D longitudinal expansion of the QGP \cite{Bjorken2}, 
which gives $T_{0}^{3}\tau_{0}=T^{3}\tau$. We take $\tau_{0}=0.5$ fm.
For simplicity we neglect variation of the initial temperature $T_{0}$ in the 
transverse directions.
We evaluated $T_{0}$ using the data on the charged hadron multiplicity 
pseudorapidity density $dN_{ch}/d\eta$ \cite{CMS_Nch,ALICE_Nch} 
and the entropy/multiplicity ratio
$dS/dy{\Big/}dN_{ch}/d\eta\approx 7.67$ obtained in \cite{BM-entropy}.
It gives $T_{0}\approx 420$ MeV for central
$Pb+Pb$ collisions at $\sqrt{s}=2.76$ TeV.
For each jet we calculate accurately 
the fast parton path length in the QGP, $L$.
To take into account the fact that at times about $1-2$ units of 
the nucleus radius the QGP should cool quickly due to transverse expansion
\cite{Bjorken2}, we  
impose the condition $L< L_{max}$. We performed the computations for 
$L_{max}=8$. The bigger value  $L_{max}=10$ fm gives almost the same.

In Fig.~1 we compare the theoretical $R_{AA}$ for charged hadrons
obtained for $\alpha_{s}^{fr}=0.5$ and 0.4
to the data from 
CMS \cite{CMS_RAAch} and 
ALICE \cite{ALICE_RAAch} for  0-5\% central $Pb+Pb$ 
collisions at $\sqrt{s}=2.76$ TeV.
The results are presented for the radiative mechanism alone
and with the collisional energy loss.
We show our results for $p_{T}\gsim 5$ GeV since at smaller momenta 
our perturbative treatment is hardly applicable.
Fig.~1 shows that the collisional mechanism suppresses $R_{AA}$ by
$\sim 20$\% at $p_{T}\sim 10$ GeV, and $\sim 10$\% at $p_{T}\sim 100$ GeV. 
One sees that the teoretical $R_{AA}$ 
(for radiative plus collisional energy loss)  for the window
$\alpha_{s}^{fr}\sim 0.4-0.5$ 
agrees reasonably with the experimental data. The agreement is somewhat better
for $\alpha_{s}^{fr}=0.4$.
In Fig.~2 we compare our results 
with the ALICE data 
\cite{ALICE_RAA_D1,ALICE_RAA_D2} on the $R_{AA}$ for $D$-mesons in 
$Pb+Pb$  collisions 
at $\sqrt{s}=2.76$ TeV for 0-20\% and 0-7.5\% centrality bins.
Fig.~2 shows the results for the $c\to D$ fragmentation.
We found that the effect of the $b$-quark (due to $b\to B\to D$ vacuum
fragmentation) increases the $R_{AA}$ only by about 2\%.
From Fig.~2 we can conlude 
that the same window in $\alpha_{s}^{fr}$ as for
light hadrons allows  to obtaind a fairly reasonable description
of the $D$-meson data as well.

In Fig.~3 we compare our calculations of the $R_{AA}$ for non-photonic
electrons 
with the recent ALICE measurement \cite{ALICE_RAAe}. We show the contibution
from the charm and bottom quarks separately and the total electron $R_{AA}$.
Note that for the bottom 
quark  our treatment of the collisional mechanism as a pertubation
to the radiative one, with the help of (\ref{eq:50}), loses accuracy 
at $p_{T}\lsim 5-6$ GeV. In this region the collisional correction 
becomes too large for the predictions to be robust. It happens since  
the $R_{AA}$ becomes sensitive to the low energy region where for the bottom quark
$\Delta E_{col}\gsim \Delta E_{rad}$. Evidently, in this 
regime the 
radiative and collisional
mechanisms must be treated on an even footing. Unfortunately, this 
problem remains unsolved.
For the charm quark this complication does not arise since 
accross the whole energy range the collisional
energy loss remains relatively small \cite{Z_Ecoll}.
Fig.~3 shows that at $p_{T}\gsim 6-7$ GeV our results 
agree with the data fairly well.
Note that for the RHIC conditions our results also agree
reasonably with the data.
For the 0-5\% central $Au+Au$ collisions at $\sqrt{s}=200$ GeV 
for $\alpha_{s}^{fr}\sim 0.5-0.6$
(what is needed for agreement with the $R_{AA}$ for pions)
at $p_{T}\sim 6-8$ GeV our calculations give the electron 
$R_{AA}\sim 0.25-0.35$, 
which agrees reasonably
with the STAR \cite{STAR_e} measurement. 
A detailed discusion of the non-photonic electrons
for RHIC and LHC energies will be given in a forthcoming publication.

Thus, our pQCD model with the radiative energy loss combined
with relatively small collisional energy loss 
gives a reasonable description of the latest LHC data on the $R_{AA}$
both for light and heavy flavors at $p_{T}\gsim 5$ GeV.

\vspace{.2cm}
\noindent {\bf 4}. 
In summary, we have examined the flavor dependence of the nuclear modification
factor $R_{AA}$ in the pQCD picture and checked its consistency with
that observed at LHC.
We show that the LHC data on the $R_{AA}$ 
for charged hadrons \cite{CMS_RAAch,ALICE_RAAch} and $D$-mesons
\cite{ALICE_RAA_D1,ALICE_RAA_D2} 
in central $Pb+Pb$ collisions at $\sqrt{s}=2.76$ TeV
can be reasonably described in the pQCD scheme,
universal for light and heavy flavors with relatively small 
collisional energy loss.
We found that the ALICE data \cite{ALICE_RAAe} on the $R_{AA}$ 
for non-photonic electrons can be described fairly well in our model as well.

We conclude that the recent LHC data on the $R_{AA}$ for the light and 
heavy jets
give strong support for the validity of the pQCD parton mass dependence of 
the energy loss with relatively small effect of the collisional mechanism.
The collisional mechanism becomes very important only for the bottom quark
at momenta $\lsim 6-8$ GeV. For accurate pQCD calculations in this region
a better understanding of the interplay 
of the radiative and collisional mechanism is required.

\vspace {.7 cm}
\noindent
{\large\bf Acknowledgements}

\noindent
I am grateful to the ALICE Collaboration 
for providing me with the 
ALICE data shown in Fig.~1.  
This work is supported 
in part by the 
grant RFBR
12-02-00063-a and the program
SS-6501.2010.2.

\vskip .5 true cm

\newpage
\begin{figure} [t]
\vspace{.7cm}
\begin{center}
\epsfig{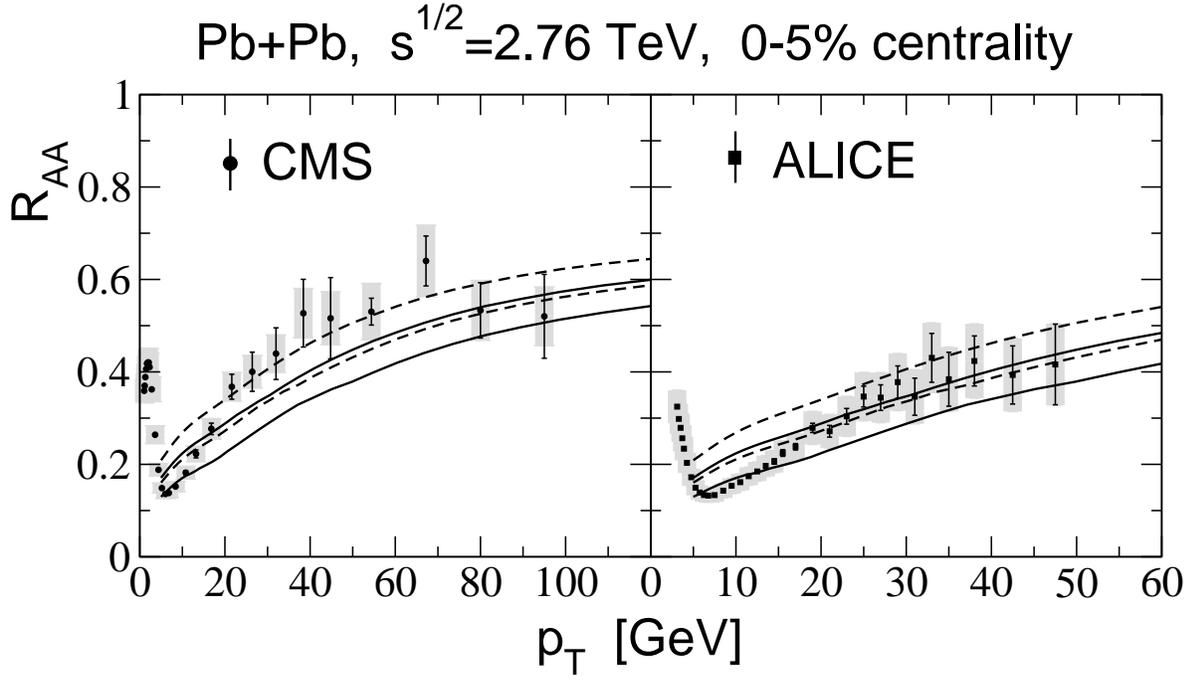}
\end{center}
\caption[.]
{
The nuclear modification factor for charged hadrons at $y=0$
for 0-5\% central $Pb+Pb$ collisions
at $\sqrt{s}=2.76$ TeV for $\alpha_{s}^{fr}=0.4$ (upper curves) and 
0.5 (lower curves).
The solid line shows the calculations with the radiative 
and collisional energy loss, and
the dashed line shows the results for
the radiative mechanism alone.
The experimental points are the data from 
CMS \cite{CMS_RAAch} (circles) and ALICE \cite{ALICE_RAAch} (squares).
Systematic errors
are shown as shaded areas. 
}
\end{figure}

\begin{figure} [t]
\begin{center}
\epsfig{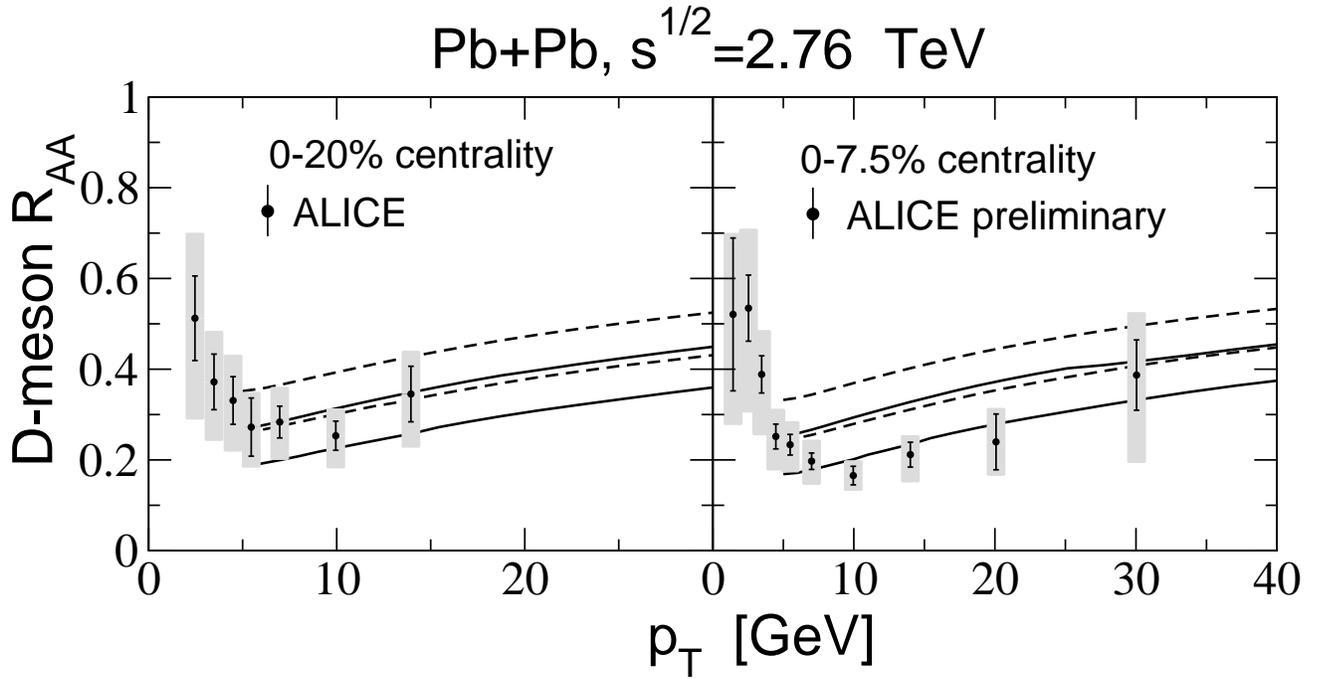}
\end{center}
\caption[.]
{
The $D$-meson nuclear modification factor 
for 0-20\% (left) and 0-7.5\% (right) central 
$Pb+Pb$ collisions at $\sqrt{s}=2.76$ TeV for $\alpha_{s}^{fr}=0.4$
(upper curves) and 0.5 (lower curves) at $y=0$.
The solid line shows the calculations with the radiative 
and collisional energy loss, and
the dashed line shows the results for
the radiative mechanism alone.
The experimental points are the ALICE data \cite{ALICE_RAA_D1} (left panel),
\cite{ALICE_RAA_D2} (right panel)
for average $D^{0}$,$D^{+}$,$D^{^{*}+}$.
Systematic errors
are shown as shaded areas. 
}
\end{figure}

\begin{figure} [t]
\begin{center}
\epsfig{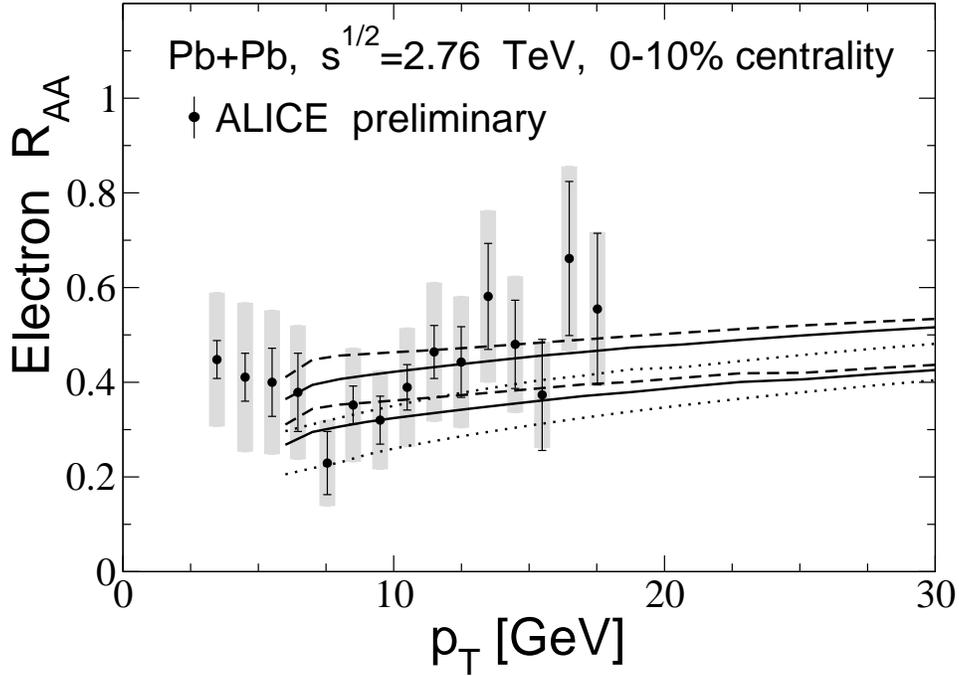}
\end{center}
\caption[.]
{
The electron nuclear modification factor for
0-10\% central 
$Pb+Pb$ collisions at $\sqrt{s}=2.76$ TeV for $\alpha_{s}^{fr}=0.4$
(upper curves) and 0.5 (lower curves) at $y=0$.
The solid line shows the total $R_{AA}$, the dotted and dashed lines show the 
$R_{AA}$ for charm and bottom contributions, respectively.
The experimental points are the preliminary ALICE data \cite{ALICE_RAAe}.
Systematic errors
are shown as shaded areas. 
}
\end{figure}

\end{document}